\documentstyle[12pt,aas2pp4,epsfig]{article}

\def\bs{\bigskip}
\def\be{\begin{equation}}
\def\ee{\end{equation}}
\def\ea{{\it et al.}\,}

\def\eg{{\it e.g.},\,}

\def\rel{relativistic \,}

\begin{document}

\hfill\today
\title{Spectral Analysis of RXTE Observations of A3667}

\author{Yoel Rephaeli\altaffilmark{1,2}, and Duane 
Gruber\altaffilmark{3}}

\affil{$^1$Center for Astrophysics and Space Sciences, 
University  of California, San Diego,  La Jolla, CA\,92093-0424}

\affil{$^2$School of Physics and Astronomy, 
Tel Aviv University, Tel Aviv, 69978, Israel}

\affil{$^3$4789 Panorama Drive, San Diego CA 92116}

\begin{abstract}

X-ray emission from the cluster of galaxies A3667 was measured by the 
PCA and HEXTE experiments aboard the RXTE satellite during the period 
December 2001 - July 2002. Analysis of the $\sim 141$ ks RXTE 
observation and lower energy ASCA/GIS data, yields only marginal 
evidence for a secondary power-law emission component in the 
spectrum. The 90\% confidence upper limit on nonthermal emission 
in the 15-35 keV band is determined to be $2.6 \times 10^{-12}$ 
erg-cm$^{-2}$ s$^{-1}$. When combined with the measured radio flux 
and spectral index of the dominant region of extended radio emission, 
this upper limit implies a lower limit of $\sim 0.4\, \mu$ G on the 
mean, volume-averaged, intracluster magnetic field in A3667.

\end{abstract}


\section{Introduction} 
The spectral and spatial resolutions of current X-ray satellites 
allow a more realistic description of gas properties and a search 
for new phenomena in clusters of galaxies. Spectral diagnostics 
alone can reveal the presence of a second spectral component that 
may indicate either a more complex temperature distribution than a 
simple isothermal (with its associated primary thermal emission), 
or an appreciable energetic electron population that radiates 
non-thermally. Clearly determining that the gas is non-isothermal 
is important for a more precise description of IC gas properties, 
evolution, and for estimation of the cluster total mass, as well 
as for our ability to use the gas as a more precise cosmological 
probe. The detection of an appreciable level of non-thermal (NT) 
emission in clusters with measured extended regions of radio 
emission (\eg, Rephaeli 1977, 1979) is also very important as an 
essential second observable which is needed in order to determine 
the strength of magnetic fields, and for estimating densities and 
energy content of \rel electrons (and protons). 

While considerable work is being done to investigate the temperature 
structure of intracluster gas, relatively few searches for non-thermal 
(NT) emission were carried out. These began with archivel analysis of 
HEAO-1 data (Rephaeli, Gruber \& Rothschild 1987, Rephaeli \& Gruber 1988), 
and continued with CGRO (Rephaeli, Ulmer \& Gruber 1994) and ASCA 
(Henriksen 1999) observations, yielding only upper limits on spectral 
power-law components. The improved sensitivity and wide spectral band 
of the RXTE and BeppoSAX allowed a more detailed spectral analysis of 
long exposure measurements that resulted in significant evidence for 
NT emission in Coma (Rephaeli, Gruber \& Blanco 1999, Fusco-Femiano \ea 
1999, Rephaeli \& Gruber 2002), A2256 (Fusco-Femiano \ea 2000, Rephaeli \& 
Gruber 2003), A2319 (Gruber \& Rephaeli 2002), A119 \& A754 (Fusco-Femiano 
\ea 2003a), and in the moderately distant cluster RXJ0658-5557 (Petrosian 
2003). However, the claimed NT emission in A754 could possibly be from 
a radio galaxy (Henriksen, Hudson \& Tittley 2003).
    
Results of two different analyses of a second BeppoSAX observation of 
Coma were reported very recently: According to Rossetti \& Molendi (2003), 
the full PDS dataset (with a total on-source exposure of $\sim 166$ ks) 
no longer shows significant evidence for a NT component, a claim that 
is disputed by Fusco-Femiano \ea (2003b). Analysis of the same data by 
the latter authors yields a very significant NT compoent at a level 
that is only slightly lower than originally reported by Fusco-Femiano \ea 
(1999). 

Since NT electron populations have a wide energy range, their emission 
could possibly be detected also in the EUV range. It has been claimed 
that low energy emission observed in a few clusters by ROSAT and,  
in particular, by the EUVE (in the $65-245$ eV band), is in excess of 
what is predicted from thermal emission by IC gas, and that the 
excess emission is NT (\eg, Lieu \ea 1996, Sarazin \& Lieu 1998, Bowyer 
\ea 1999, 2003). Due to the very limited spectral range of the 
EUVE measurements this identification is uncertain. Analysis of line 
and continuum XMM measurements of soft emission from a sample of 
clusters seems to suggest that the excess emission is thermal emission 
from warm gas (Kaastra \ea 2003).

The number of clusters observed at high ($>30$ keV) energies is only a 
small fraction of the $\sim 40$ clusters in which extended radio emission 
has already been measured. It is of considerable interest to enlarge the 
small sample in order to begin a more systematic study of NT phenomena in 
clusters. Here we report the results of an analysis of $\sim 141$ ks RXTE 
observation of the `merging' cluster A3667.

\section{Observations and Spectral Analysis}

A3667, a rich southern cluster at $z=0.055$, has a bimodal galaxy 
distribution, large velocity dispersion, distorted X-ray 
morphology, and complex extended regions of radio emission 
(\eg, Rottgering \ea 1997). These features are thought to 
indicate that the cluster is undergoing strong merging activity. 
Of particular relevance is the very large, elongated and off-center 
region of radio emission. The spectrum of this dominant source 
(point sources are estimated to contribute only a few percent of 
the total emission) can be described in terms of an overall, single 
value of the spectral (energy) index, $\alpha \sim 1.1$, 
in the frequency range $\sim 0.4 -2.3$ GHz, and a flux of 
$5.5 \pm 0.5$ Jy at 843 MHz (Rottgering \ea 1997).

Analysis of ASCA observations of A3667 yielded a mean gas 
temperature of $7.0 \pm 0.6$ keV over a large region with $\sim 22'$ 
radial extent (Markevitch \ea 1999). More recently {\it Chandra} 
measurements have led to a more detailed temperature and brightness 
structure in the central region showing evidence for large 
temperature gradients, including a region where the gas temperature 
is $\sim 11$ keV, possibly due heating of the gas merger shocks 
(Vikhlinin \ea 2001). 

A3667 was observed with RXTE for $\sim 167$ ks during the period 
December 2001 - July 2002. The application of data selection criteria 
recommended by the RXTE project results in 141 ks of screened PCA 
data, spaced irregularly over the observing period. These were 
collected in two of the 5 detectors. For the HEXTE, which 
beam-switches observations with 32-second dwells between source and 
background fields, and has in addition about 50\% detector dead time, 
the net observation time was 54.6 ks with each of the two clusters. 
A systematic error of 0.8\% per energy channel was added in 
quadrature to the statistical error of the PCA data; no systematic 
error was used with HEXTE data. On time scales of two weeks or 
longer, the limit to variability observed with PCA was less than 
1\%. Because of the much lower signal to background, corresponding 
HEXTE limits to variability are weaker, about 20\%. 

To extend the spectral range lower, analysis was conducted jointly with 
archival ASCA GIS observation (of 1995 April 16) lasting and 39 ks. GIS2 
and GIS3 0.8 -- 8.0 keV spectra were accumulated in a field with a 
diameter of 14 arcminutes centered on the cluster, a region that included 
the great majority of the cluster emission. SIS data for the observation 
were found unsuitable for analysis because of noise.  Systematic errors 
were not added to the GIS data.   

Preliminary spectral analysis on the direct instrument data showed
no features which departed on small scales (i.e. spectral lines or 
edges) from the thermal form which clearly dominates the bulk of the 
observed cluster spectrum. We therefore found it appropriate to 
combine data in order to make small differences of $\chi-^{2}$ from 
model fitting more distinguishable against the statistical noise of 
the highly oversampled data. As a first step we combined the data 
from the two PCA detectors and also the data of the two HEXTE 
clusters. For all three instruments we then joined the counts in 
adjacent energy channels into wider energy bins at a density of two 
to three per detector energy resolution element. Doing so reduced 
the number of energy bins in the analysis from over one thousand to 
60. These channels are displayed in the figure. It should be noted 
that confidence intervals for parameter estimates, as shown in the 
Table, depend on differences of $\chi-^{2}$, and are only weakly
sensitive to choices of binning.

Both the RXTE and ASCA data are collected essentially from the entire 
cluster, so that the joint fits here need not consider possible spectral 
differences resulting from gradients in the cluster. To account for possible 
inter-calibration errors between the instruments, adjustable scaling constants 
are employed.  These are treated as `uninteresting' parameters in the 
fitting, provided (as is the case here) that they float to values within 
5 -- 15\% of unity, consistent with our experience in analyzing other sources 
with this combination of instruments. 
 
\begin{figure}[t]
\centerline{\psfig{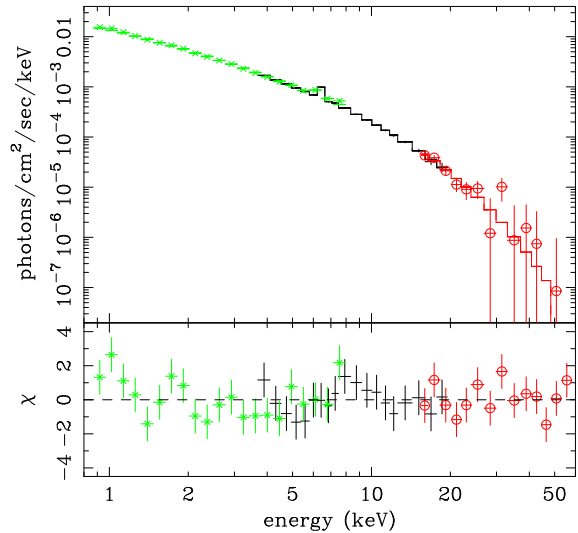}}
\figcaption{RXTE and ASCA/GIS spectrum of the A3667 with folded
Raymond-Smith ($kT \simeq 7.3$) model. ASCA/GIS data are shown in
(green) asterisks, (black) crosses are PCA data, and HEXTE data
points are marked with (red) circles (with 68\% error bars). The
total fitted spectrum is shown with a histogram, and deviations
between the model and data points (normalized to the respective
standard error) are shown in the lower panel.}
\end{figure}

We have considered three spectral models: thermal emission from 
isothermal gas (based on a Raymond-Smith emission code), two-temperature 
thermal, and a thermal plus a power-law model. The hydrogen column was 
fixed at the Galactic value, $N_H \simeq 4.7 \times 10^{20}$ cm$^{-2}$. 
With the lower threshold at 0.8 keV, the GIS data are insensitive to the 
value of $N_H$ unless it is much larger. Indeed, if this value is allowed 
to float in the fits a best-fit is found which is only slightly lower 
than our assumed value. Using these models, fits to the joint data 
provide only weak evidence of the need for an extra component beyond 
isothermal. The $\chi^2$ of 56.4 (57 degrees of freedom [dof]) for an 
isothermal fit is acceptable. Inclusion of a second component 
reduces $\chi^2$ modestly to 46.6 (55 dof) with an extra 0.9 keV thermal 
component, or to 51.3 (56 dof) with an extra power-law of index 2.1, as 
determined from the radio spectrum. If the power law index is allowed to 
vary, its best-fit photon index assumes the rather steep value of 5.0, 
but it is almost unconstrained. With four ``interesting'' parameters, 
-- temperature, abundance, and normalization of the thermal component, 
plus power-law normalization -- the change in $\chi^2$ (Lampton \ea 1976) 
gives 90\% error limits for the 2-20 keV power-law flux of 
$(0.0-8.1)\times 10^{-11}$ erg-cm$^{-2}$ s$^{-1}$. The temperature for 
the main spectral component is in the range $\sim 7.3-7.5$ keV for all 
three cases, with formal ($1\sigma$) errors of $\sim 0.1-0.3$ keV. 

This RXTE observation yields evidence for a second spectral component 
which is just significant at the 90\% confidence level. With the 
spectral photon index fixed at 2.1, the value predicted from the 
radio spectrum, the best-fit 2-20 keV flux is $(4.0 \pm 1.8)\times 
10^{-12}$ erg-cm$^{-2}$ s$^{-1}$. The 90\% confidence upper limit on 
the 15-35 keV flux is $2.6 \times 10^{-12}$ erg-cm$^{-2}$ s$^{-1}$ 
(with four relevant parameters). The fraction of the total flux in the 
best-fit secondary component, whether low-temperature thermal or power-law, 
is small, of the order of a few percent. We report these values, together 
with 90\% confidence errors, as ``secondary flux fraction'' in the Table 
for three interesting energy bands. Nevertheless, given the possible 
effects of background subtraction and calibration errors, we feel that 
we cannot claim detection of a second component at 90\% confidence, even 
though this is a formal result of the analysis.                     

\begin{table*}[t]        
\caption{Results of the spectral analysis}
\bs

\begin{tabular}{|l|ccc|} 
\hline 
Parameter & Single Thermal & Double Thermal & Thermal + Power-law \\  
\hline   
$kT_1$ (\rm{keV}) &$7.3\pm 0.2$ & $7.5^{+0.7}_{-0.3}$ & $7.3\pm 0.2$ \\      
                  &            &                        &                  \\
$kT_2$ (\rm{keV}) &            & $0.9^{+3.7}_{-0.8}$ &                  \\
                  &            &                        &                  \\            
Secondary flux fraction &      &                        &                      \\       

%\,\,\, 0.5-2 keV   &           & $0.056^{+0.158}_{-0.056}$  & $0.079^{+0.085}_{-0.079}$   \\
\,\,\, 2-10 keV    &           & $0.003^{+0.155}_{-0.003}$  & $0.045^{+0.048}_{-0.045}$      \\
\,\,\, 0.8-40 keV &         & $0.016^{+0.153}_{-0.016}$ &$0.052^{+0.056}_{-0.052}$ \\   
                  &               &                        &                  \\       
Abundance (solar)  & $0.22 \pm 0.04$ & $0.22\pm 0.05$ & $0.24\pm 0.05$  \\                
\hline
\end{tabular} 
         
\tablenotetext{}{All quoted errors are at the 90\% confidence level. } 
\end{table*}               

A3667 was observed by BeppoSAX for $\sim 113$ ks in May 1998 and 
October 1999. Analysis of the measurements (Fusco-Femiano \ea 2001) 
showed marginal evidence ({\it formally} significant at the $\sim
2.6\sigma$ level) for a secondary power-law component, and this only 
if the gas temperature was fixed at the value ($7$ keV) determined 
from previous ASCA measurements. Clearly, the need to {\it assume} a 
value for the temperature, rather than determining it in a
simultaneous fit to the parameters of both components (a procedure 
that is difficult to accomplish with BeppoSAX due to the lack of 
spectral overlap between the PDS and the lower energy MECS
experiments), introduced a substantial uncertainty in the deduced 
significance of any power-law emission. Therefore, only an upper limit
on nonthermal emission was reported by Fusco-Femiano \ea (2001); in 
the 15-35 keV range the flux limit is $4.2\times 10^{-12}$ 
ergs cm$^{-2}$s$^{-1}$, moderately higher than our limit in the same 
energy band. 
 
\section{Discussion}

A3667 is the fourth (following Coma, A2319 \& A2256) in a sample of 
clusters with extended radio emission whose RXTE observations were 
analyzed by us, and the only one for which we do not find clear 
evidence for a second spectral component. With no spatial 
information, the absence of evidence for a secondary thermal emission 
component does not yield useful information on a large scale 
temperature gradient. In a cluster with extended radio emission, the 
main interest in obtaining an upper limit on NT emission stems from 
the fact that it sets a lower limit on the mean, volume-averaged 
value of the magnetic field in the central region of the cluster. 
With the above value of the flux upper limit, and a spectral 
index of $2.1$ inferred from the radio spectrum, we set a lower 
limit of $\sim 0.4$ $\mu$G on the mean value of the magnetic field. 

In assessing the meaning of this limit it has to be realized that 
radio emission in A3667 has a complex morphology, with very 
different field values in different regions. From a more basic 
point of view, it should also be remembered that several implicit 
assumptions are usually made in linking the synchrotron and Compton 
formulae in order to determine $B_{rx}$ from radio and X-ray 
measurements (Rephaeli 1979, Goldshmidt \& Rephaeli 1993). These 
are too often ignored, especially when values of $B_{rx}$ are 
contrasted with field values deduced from co-added statistical 
analysis of Faraday rotation measurements (\eg, Clarke, Kronberg, 
and B\"ohringer 2001). The latter method is also prone to substantial 
inherent (\eg, Newman, Newman \& Rephaeli 2002) and systematic 
(Rudnick \& Blundell 2003, but see Ensslin \ea 2003) uncertainties.

Since the number of clusters observed at high X-ray energies is small, 
the observation of additional clusters (either with or without known 
extended radio emission) is of obvious interest, even if only an upper 
limit is obtained on NT emission: While observations of more clusters 
with extended radio emission are very desirable, observations of other 
clusters, interesting in their own right, are also useful as a control 
sample.

\acknowledgments
We thank the referee for useful comments. This project has been 
supported by a NASA grant at UCSD.

\parskip=0.015in
\def\ref{\par\noindent\hangindent 20pt}

\end{document}